\documentclass[
aps,
pra,
12pt,
reprint,
longbibliography,
floatfix
]{revtex4-2}

\usepackage[utf8]{inputenc}
\usepackage[T1]{fontenc}
\usepackage[english]{babel}
\usepackage{amsmath}
\usepackage{physics}
\usepackage{graphicx}
\usepackage[colorlinks=true, allcolors=black]{hyperref}

\usepackage{xcolor}
\usepackage{graphicx}
\usepackage[left=23mm,right=13mm,top=35mm,columnsep=15pt]{geometry} 

\usepackage{placeins}
\usepackage[T1]{fontenc}
\usepackage{appendix}

\usepackage{comment}

\usepackage{color}

\usepackage{soul}

\AtBeginDocument{}
\AtBeginDocument{}

\begin{document}
	\title{Non-adiabatic couplings as a stabilization mechanism in long-range Rydberg molecules}
 \author{Aileen A. T. Durst}
 \email[These authors contributed equally.]{}
	\affiliation{Max Planck Institute for the Physics of Complex Systems,  Nöthnitzer Str. 38, 01187 Dresden, Germany}
 \author{Milena Simi\' c$^*$}
 \email[Correspondence should be addressed to ]{msimic@pks.mpg.de}
	\affiliation{Max Planck Institute for the Physics of Complex Systems,  Nöthnitzer Str. 38, 01187 Dresden, Germany}
 \author{Neethu Abraham}
	\affiliation{Max Planck Institute for the Physics of Complex Systems,  Nöthnitzer Str. 38, 01187 Dresden, Germany}
  \author{Matthew T. Eiles}
	\affiliation{Max Planck Institute for the Physics of Complex Systems,  Nöthnitzer Str. 38, 01187 Dresden, Germany}
	\date{\today} 
\begin{abstract}

Long-range Rydberg molecules are typically bound in wells formed in their oscillatory potential energy curves. In alkaline Rydberg molecules, bound vibrational states exist even when these potential wells are disrupted by level repulsion from the steep butterfly potential energy curve induced by a scattering shape resonance. The binding in this case is attributed to quantum reflection. However, the rapidly varying regions of the potential energy landscape where quantum reflection occurs often coincide with regions where non-adiabatic coupling becomes significant. 
By comparing the molecular states calculated within the Born-Oppenheimer approximation, where quantum reflection is the only binding mechanism, with those obtained from the full set of coupled channel equations, we can assess the effects of non-adiabatic coupling on vibrational energies and lifetimes.
Our findings show that these couplings can stabilize the molecule by providing an additional barrier which protects the vibrational states from predissociation and non-radiative transitions. There can also be extreme cases where non-adiabatic coupling completely dominates the binding and the molecular lifetimes saturate at the atomic Rydberg lifetime.
\end{abstract}

\maketitle

\section{Introduction}
Ultracold chemistry has recently developed into a dynamic field marked by a strong synergy between theory and experiment and by rapid advancements in the coherent control of atoms, ions, and molecules \cite{karman_ultracold_2024, langen_quantum_2024}. 
Such advancements include the ability to build molecules atom-by-atom in optical tweezers \cite{ruttley_formation_2023, zhang_forming_2020}, the simultaneous cooling \cite{ravi_cooling_2012, vuletic_laser_2000} and trapping of ions together with atoms \cite{ewald_observation_2019, hirzler_observation_2022, haerter_cold_2014,deis_cold_2024} and the control of their interactions with Feshbach resonances \cite{chin_feshbach_2010, kohler_production_2006,weckesser_observation_2021}, the microwave shielding of ultracold molecules to dramatically reduce chemical reactivity and four-body loss \cite{andereg_observation, bigagli_observation_2024}, and the use of Rydberg atoms to facilitate the creation of ultracold ions \cite{engel_observation_2018, kleinbach_ionic_2018} and to form long-range Rydberg molecules \cite{greene_creation_2000,bendkowsky_observation_2009, niederprum_observation_2016}. 
These molecules, which form when a Rydberg atom with principal quantum number $n$ binds to a ground-state atom through the elastic scattering of the Rydberg electron off of the ground-state atom, are a particularly dramatic example of the chemistry accessible at ultracold temperatures \cite{eiles_trilobites_2019,fey_ultralong-range_2020,shaffer_ultracold_2018,sasmannshausen_experimental_2015}. 
They are very fragile, with binding energies on the sub-mK scale and lifetimes limited to that of the Rydberg atom \cite{whalen_lifetimes_2017,butscher_lifetimes_2011}.
Nevertheless, their large bond lengths and dipole moments, in excess of several hundreds of nanometers and a few kiloDebye, respectively, make them fascinating objects with which to study ultracold chemical processes and few-body physics on a grand scale \cite{karman_ultracold_2024, schlagmuller_ultracold_2016}. 

\begin{figure}
    \centering
    \includegraphics[width= 0.45\textwidth]{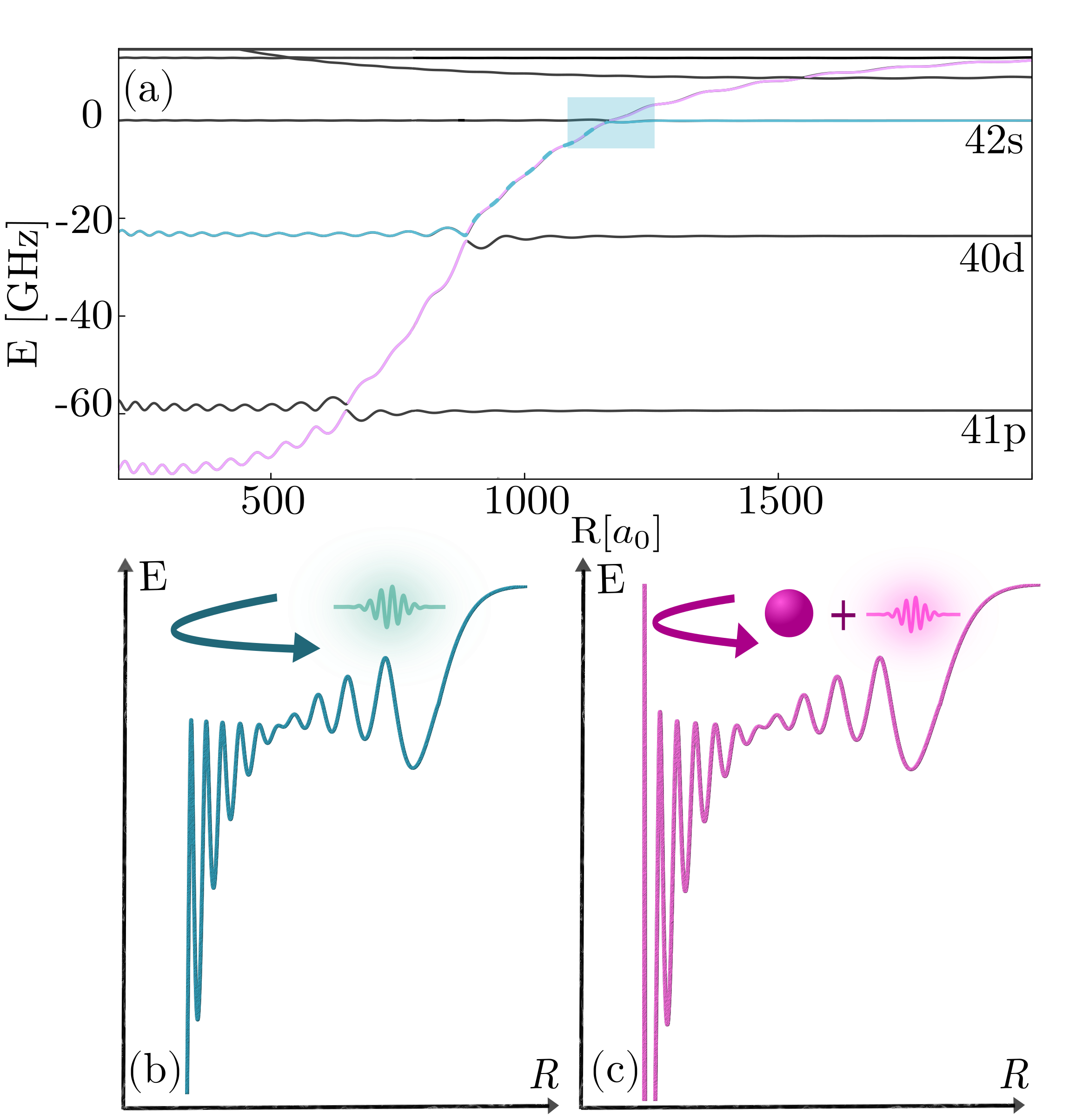}
    \caption{a) Adiabatic potential energy curves of a Rb long-range Rydberg molecule resulting from $S$- and $P$-wave scattering of the Rydberg electron off of the ground-state atom. The butterfly potential energy curve, whose rapid energy variation is a consequence of the $P$-wave shape resonance, is shown in pink. It descends steeply from the degenerate manifold of hydrogenic states. The avoided crossing studied in this article is highlighted in the shaded region. The Born-Oppenheimer potential energy curve utilized in all single channel calculations is colored in blue.  
    The sketches (b) and (c) emphasize the difference between the two binding mechanisms -- pure quantum reflection (b) and quantum reflection enhanced by non-adiabatic couplings (c). The molecular states (portrayed as a wave packet) in the Born-Oppenheimer (turquoise) potential curve are bound solely by the steep drop of the potential, while the Born-Huang (magenta) potential curve exhibits a sharp barrier peak arising due to non-adiabatic effects.  }
    \label{fig:potentials}
\end{figure}

One remarkable feature of long-range Rydberg molecules is the “butterfly” potential energy curve, which plunges steeply down from one degenerate Rydberg manifold with principal quantum number $n$ before eventually settling down above the adjacent $n-1$ Rydberg manifold \cite{hamilton_shape-resonance-induced_2002, chibisov_energies_2002,khuskivadze_adiabatic_2002}. 
This phenomenon is observed when the ground state atom- electron scattering possesses a $P$-wave shape resonance -- a condition met in all alkaline atoms \cite{bahrim_negative-ion_2001,khuskivadze_adiabatic_2002,bartschat_ultralow-energy_2003} as well as several other species \cite{hutt_phase_1975, bartschat_convergence_2004}.
Whenever this curve approaches one of the nearly flat potential curves associated with Rydberg states having low angular momentum $\ell<3$ and sizable quantum defects, the two curves repel each other, forming an avoided crossing. 
This leads to the collection of Born-Oppenheimer potential energy curves (PECs) seen in \autoref{fig:potentials}(a).
The repulsion between adiabatic curves causes the long-range potentials to lack an inner barrier (see the isolated potential curve shown in \autoref{fig:potentials}(b)) which would be able to confine the vibrational states. 
This should make these states susceptible to rapid decay, as $\ell$-changing collisions and chemi-ionization occurring at small internuclear distance $R$ will reduce the molecule's lifetime \cite{geppert_diffusive-like_2021, althon_exploring_2023, schlagmuller_ultracold_2016,niederprum_giant_2015}.

Nevertheless, the literature on Rydberg molecules abounds with observations of relatively stable vibrational states, even in extreme cases where the outermost potential well is disrupted by the butterfly curve \cite{bendkowsky_rydberg_2010, butscher_lifetimes_2011,peper_heteronuclear_2021}. 
Quantum reflection has been posited as a mechanism to suppress the decay of the vibrational states and stabilize the molecules: perhaps counterintuitively, a wave packet traveling towards smaller $R$ values ricochets off of the drop in the potential and remains bound in obvious contrast to a classical particle \cite{bendkowsky_rydberg_2010, friedrich_quantum_2002}.
In this picture, it is the rapid change in the $n\ell$ potential curve near its encounter with the butterfly curve which prevents molecules from decaying. 
The sketch in \autoref{fig:potentials}(b) illustrates this mechanism.

While this explanation qualitatively explains the existence and lifetimes of these molecules, the fact that this quantum reflection phenomenon occurs in the vicinity of a narrow avoided crossing calls into question whether or not the Born-Oppenheimer approximation is appropriate. 
In particular, since the diagonal correction to the Born-Oppenheimer approximation is always positive, it can lead to a sharp peak in the PECs (see \autoref{fig:potentials}(c)) which could also classically trap the molecule.
This leads to a competition between non-adiabatic effects and quantum reflection for the binding mechanism.

In this article, we aim to elucidate the influence of non-adiabatic effects on Rydberg molecular states. 
While previous studies of non-adiabatic effects in Rydberg molecules have focused on how they modify binding energies \cite{hummel_vibronic_2023, srikumar_nonadiabatic_2023}, we extend the focus to the stability of these molecules.
By comparing the lifetimes calculated within the Born-Oppenheimer approximation, where quantum reflection is the sole binding mechanism, to those determined by the full solution involving non-adiabatic couplings, we assess the impact of non-adiabaticity on the binding process.

\section{Theory}
The molecular wave functions in the adiabatic representation, $\Psi_\nu(\vec r, \vec R) = \sum_i\psi_i(\vec r;\vec R)\chi_{i\nu}(\vec R)$, are obtained from the solutions of the full coupled-channel equations for the nuclear degree of freedom, 
\begin{align}\label{eq:H_coupled}
    \begin{split}
    0=\left(-\frac{\nabla_R^2}{2\mu} + U_i(R) -E_\nu\right)\chi_{i\nu}(R) \\+\sum_j \Lambda_{ij}(R)\chi_{j\nu}(R).
    \end{split}
\end{align}
Here, $\mu$ is the reduced mass of the diatomic molecule, {$\chi_{i\nu}( R)$ is the nuclear wave function in the $i$th electronic state $\psi_i(\vec r;\vec R)$, $E_\nu$ is the binding energy, $\underline\Lambda$ is the non-adiabatic coupling matrix, and we have specialized to the case of zero molecular rotation. 
\autoref{eq:H_coupled} is obtained after integrating out the electronic degrees of freedom, which results in an electronic potential energy curve $U_i(R)$ for each electronic eigenstate $\psi_i(\vec r;\vec R)$ (see \autoref{fig:potentials}). 
The electronic Hamiltonian of a long-range Rydberg molecule has been discussed in detail in many references \cite{greene_creation_2000, hamilton_shape-resonance-induced_2002,eiles_trilobites_2019,shaffer_ultracold_2018}, and it has been extended to include spin-orbit coupling, non-zero hyperfine structure, as well as external fields \cite{khuskivadze_adiabatic_2002,eiles_hamiltonian_2017,anderson_angular_2014,hummel_alignment_2019, kurz_ultralong_2014}.
We focus on Rb$(ns)$+Rb$(5s)$ long-range molecules with a spin-independent approach, which has proven adequate to describe the key features observed in experiments \cite{bendkowsky_observation_2009,bendkowsky_rydberg_2010,engel_precision_2019}.
This simplification allows us to focus on the phenomenology of non-adiabatic physics in this system.

The coupling terms 
\begin{equation}
    \Lambda_{ij}(R) = -\frac{1}{2\mu}\left(2P_{ij}(R)\frac{d}{dR} + Q_{ij}(R)\right)
\end{equation} are composed of derivatives of the adiabatic electronic states, $P_{ij}=\langle \psi_i |\frac{d}{dR}|\psi_j \rangle $ and $Q_{ij}=\sum_kP_{ik}P_{kj}+\frac{dP_{ij}}{dR}$}. 
These derivatives, in turn, are inversely proportional to the energy gap $U_i(R)-U_j(R)$.
As long as the electronic states vary smoothly with $R$ and are energetically isolated, these terms remain small.
For this reason the Born-Oppenheimer approximation neglects $\underline\Lambda$ entirely, reducing the Hamiltonian to a set of decoupled single-channel equations \cite{born_oppenheimer}.

However, at narrow avoided crossings, the electronic states change rapidly with $R$, making the derivative couplings comparable to the potentials and, therefore, not negligible \cite{worth_beyond_2004,baer_beyond_2006, hummel_vibronic_2023}.
Since the solution of the full coupled-channel problem of \autoref{eq:H_coupled} is numerically expensive and, due to the off-diagonal coupling terms, difficult to interpret and visualize, the Born-Huang approximation is often employed.
In this approach, only the diagonal terms of the coupling matrix $\Lambda_{ii}$ are included \cite{Book_born_huang}. 
These provide the first order non-adiabatic correction to the pure Born-Oppenheimer potentials while still decoupling \autoref{eq:H_coupled}. 
The correction appears as an additional peak positioned where the PEC sharply bends downwards.

Since the low-$\ell$ PECs are, in general, well-separated in energy due to their differing quantum defects, only the pairs of PECs with an avoided crossing are significantly coupled. 
Therefore, it is sufficient to consider two channels in order to converge the full-coupled calculation in this context.
Furthermore, the Rb$(ns)$+Rb$(5s)$ molecules we are interested in are energetically distant from the oscillations in the adiabatic potential curve at small R, which we assume to be flat. 
As the internuclear separation decreases, the two nuclei interact, leading to associative ionization \cite{schlagmuller_ultracold_2016, niederprum_giant_2015-1}.
By treating the molecular states as inward scattering states, where all ingoing flux is lost beyond an inner boundary $R_0$, we effectively include the different decay processes without the need for their explicit calculation.

\section{Numerical methods}
With these assumptions, we systematically study the effects of non-adiabatic physics on the molecular decay utilizing two different methods to compute the ingoing scattering states.
By contrasting the wave functions, resonance energies, and lifetimes computed from these two independent methods, we can assess the accuracy of our calculations.
This is especially useful when examining small differences between resonance energies and widths. 

The first numerical approach uses Siegert states within a single-channel scattering framework \cite{siegert_derivation_1939,tolstikhin_siegert_1997,tolstikhin_siegert_1998}.
These states satisfy the usual boundary condition for bound states, $\chi(R\to\infty)=0$, and the Siegert boundary condition, $\partial \chi(R)/\partial R|_{R=R_{0}}=-ik\chi(R=R_{0})$ at the inner boundary.
Subject to this boundary condition, the Schrödinger equation becomes a quadratic eigenvalue equation whose complex eigenvalues $-i k_n$ yield resonance positions $E_{res}$ and their respective widths $\Gamma$ via the expression,
\begin{equation}
    k_{n}^{2}/(2m) = E_{res}-i\Gamma /2. 
\end{equation}
We reformulate the quadratic eigenvalue problem into a generalized eigenvalue problem \cite{tolstikhin_siegert_1998} using a $B$-spline basis of order 12 with 2000 knot points. 
We distribute these so that there are more knot points in regions of rapid fluctuation in the potentials and coupling terms than in areas of smooth variation, and have confirmed that this number of splines is sufficient to converge all results. 
In general, vibrational states with vanishing amplitude at $R_0$ are bound states characterized by $\text{Im}(k_n)>0$ and $\text{Re}(k_n)= 0$, while eigenstates with negative $E_\mathrm{res}$ and positive $\Gamma$ correspond to quasibound resonant states. 
The existence of a positive decay rate leads to line broadening described by a Lorentz or Breit-Wigner profile  \cite{siegert_derivation_1939}.

In the second numerical approach, we extract the same resonance parameters utilizing the stabilization method.
This method is based on diagonalization of the Hamiltonian \autoref{eq:H_coupled} subject to Dirichlet boundary conditions, $\chi(R_0) = \chi(\infty) = 0$, for many different inner boundary positions \cite{mandelshtam_calculation_1993}. 
We represent the Hamiltonian using the same $B$-spline basis as in the Siegert method and change $R_0$ from $R_0^\mathrm{min}=300\,a_0$ to $R_0^\mathrm{max}=450\,a_0$.

\begin{figure*}
    \centering
    \includegraphics[width=0.95\textwidth]{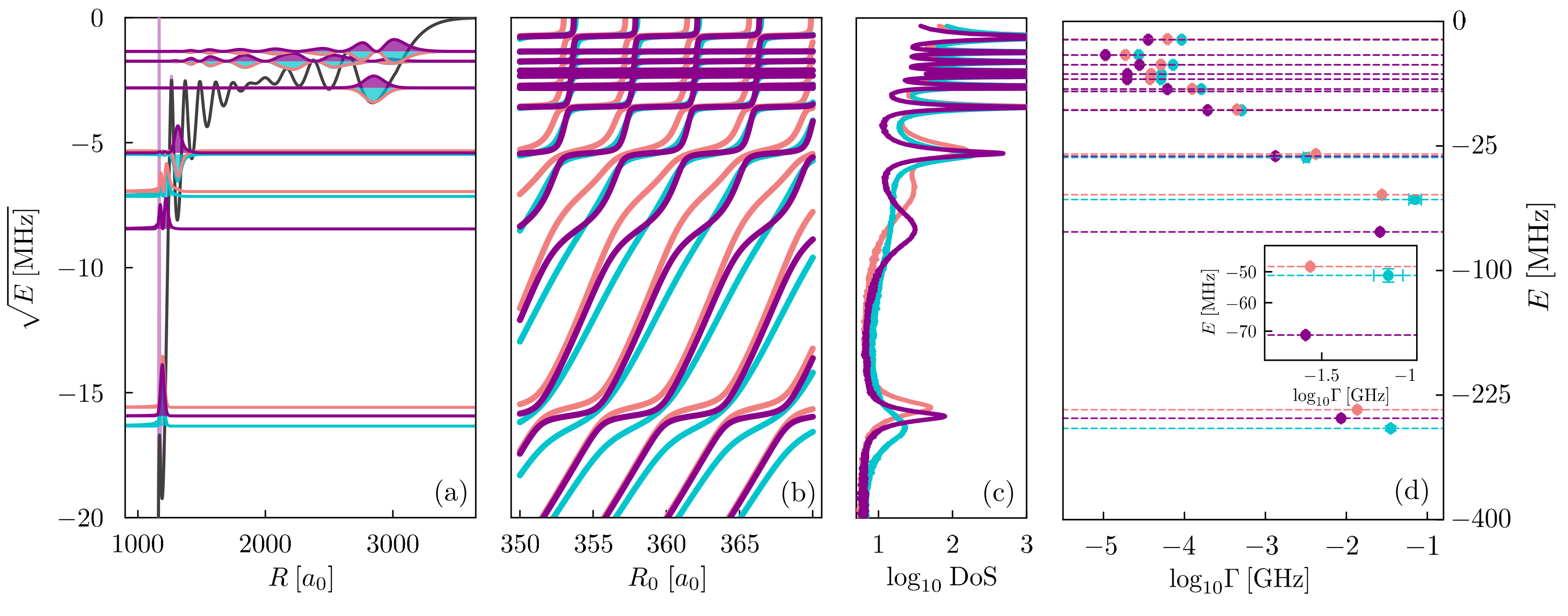}
    \caption{
    Each panel of the figure emphasizes a different aspect of the calculation, comparing Born-Oppenheimer (turquoise), Born-Huang (pink) and coupled-channel (purple) solutions.
    (a) A selection of wave functions supported by the Rb$(42s)$+Rb$(5s)$ potential energy curve. The Born-Oppenheimer potential curve is given in black, while the Born-Huang potential curve is shown in dark pink. They differ only at the avoided crossing. The energies are given with respect to the Rydberg $ns$ level. 
    (b) Stabilization diagram showing the dependence of the resonance energies on the position of the inner boundary $R_0$. The flat plateaus correspond to resonance states, showing very weak or no dependence on $R_0$. 
    (c) Density of States obtained by histogramming and normalizing the stabilization diagram. 
    (d) The widths $\Gamma$, shown on a logarithmic scale, of the corresponding vibrational states. The plot includes error bars obtained by comparing the Siegert and stabilization methods, although they are too small to be seen except in the inset. 
    Note that a square root scaling is employed to spread out the energy scale so that deeply bound vibrational states and near-threshold vibrational states can be seen on the same figure; corresponding energies are displayed on the right axis. }
   
    \label{fig1:39}
\end{figure*}

In a stabilization plot, where all eigenenergies are plotted as a function of $R_0$ (\autoref{fig1:39}(b)), it can be seen that the energies of resonant states are only weakly affected by the position of the inner boundary \cite{maier_spherical-box_1980}. 
Accordingly, by binning these energies and normalizing by the factor $dR_0/(R_0^\mathrm{min}-R_0^\mathrm{max})$, where $dR_0$ is the step size, and $[R_0^\mathrm{min},R_0^\mathrm{max}]$ is the range over which the inner boundary position is varied \cite{muller_calculation_1994},  the density of states (DoS) is obtained \cite{bressanini_generalized_2011, maier_spherical-box_1980, muller_calculation_1994, mandelshtam_calculation_1993}. 
Here, scattering resonances appear as Lorentzian peaks (see \autoref{fig1:39}(c)). 

Formally, these two methods for calculating single channel resonance profiles yield identical results. 
Differences between their results therefore provide a useful estimate of the numerical error, which allows us to distinguish meaningful physical variations in lifetimes and molecular energies between different approximations from numerical artifacts. 
We compute the relative error with respect to the Siegert values which is then displayed in all figures. 
Although generalizations of the Siegert boundary conditions to multichannel problems do exist, their implementation becomes numerically more challenging (due to the doubled matrix dimension for a fixed basis set) \cite{sitnikov_siegert_2003}.
We utilize the stabilization method alone to solve the two-channel coupled equation \autoref{eq:H_coupled}, assuming the error bar to be the same as in the Born-Huang calculation, since it already includes diagonal non-adiabatic effects. 

\section{Results}

In the following, we study the influence of non-adiabatic effects on the binding mechanism of long-range Rydberg molecules using this high precision analysis. 
First, we consider Rb$(42s)$+Rb$(5s)$ molecular states. 
The relevant Born-Oppenheimer (black) and Born-Huang PECs (light pink) are shown in \autoref{fig1:39}(a). 
Here the adiabatic potential curves repel each other in such a way that the innermost potential well of the $42s$-potential curve is situated within the steep drop of the butterfly potential curve, at an energy much deeper than the other wells. 
This well remains open towards small $R$ in the Born-Oppenheimer approximation and therefore cannot classically bind the molecule together, but nevertheless it supports molecular states (shown in blue) which rapidly diminish in amplitude as they encounter the potential valley at $R=1200 a_0$.
These states exemplify the quantum reflection binding mechanism \cite{bendkowsky_rydberg_2010}.
The vibrational states obtained using the coupled-channel equations (purple) and those obtained in the Born-Huang approximation (pink) look qualitatively similar to the Born-Oppenheimer solutions, but several of these states show sizable differences in their binding energies.

\begin{figure}[b!]
    \centering
    \includegraphics[width=0.9\linewidth,]{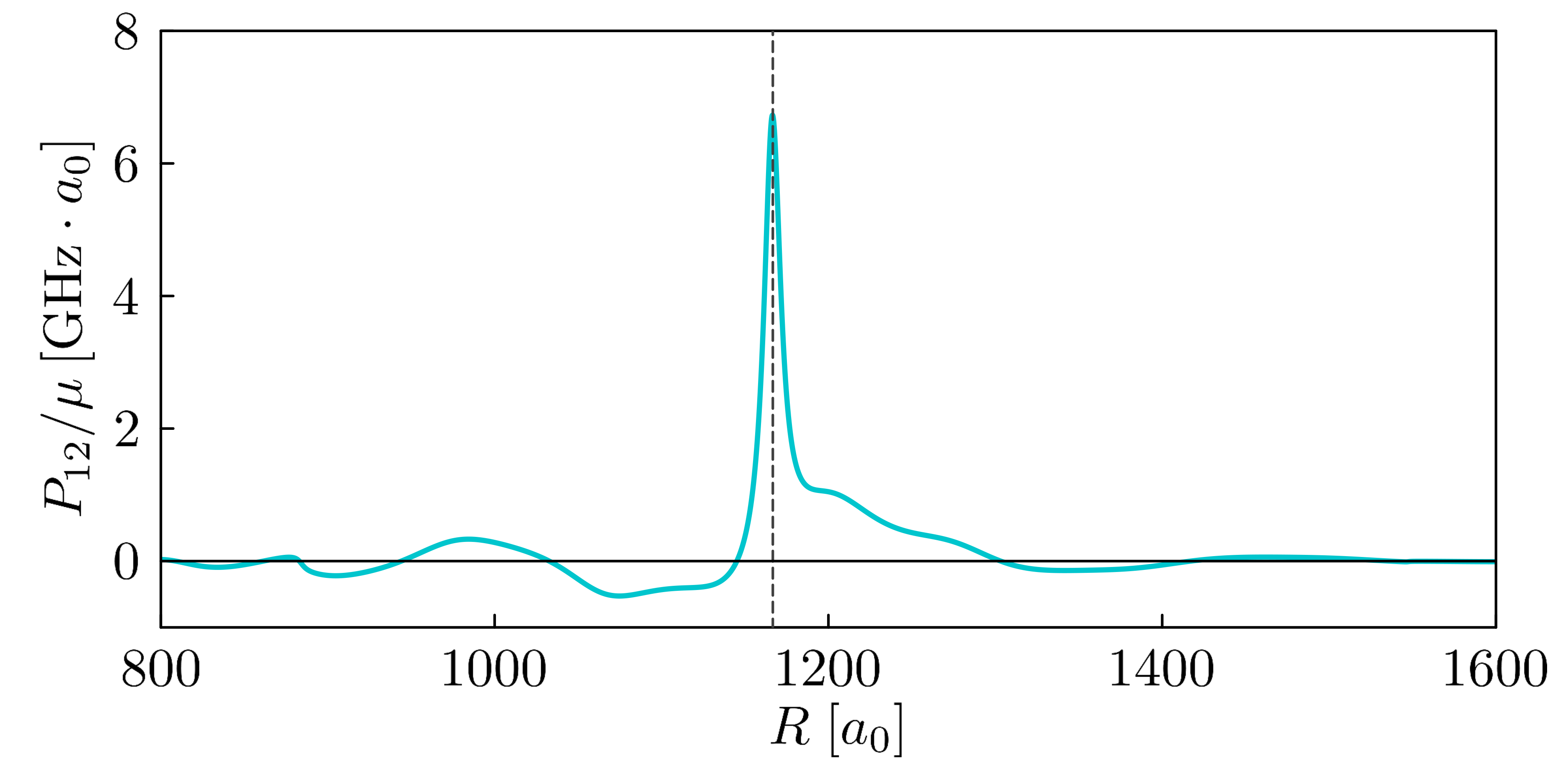}
    \caption{The off-diagonal non-adiabatic coupling matrix element P$_{12}$/$\mu$ for the Rb$(42s)$+Rb$(5s)$. The vertical dashed line indicates the position of the avoided crossing, which coincides with the maximum of $P_{12}$.}
    \label{fig:39_p}
\end{figure}

We see the strongest differences in the resonance positions in \autoref{fig1:39}(d) for the molecular states which have significant amplitude at the avoided crossing. 
This is a result of the spatial dependence of the non-adiabatic coupling elements in $\underline\Lambda$. 
These elements peak at the avoided crossing, as seen in the Born-Huang potential curve in \autoref{fig1:39}(a) and in the matrix element $P_{12}$ shown in \autoref{fig:39_p}.
Hence, the dependence of the energy shifts can be understood by considering the effect of spatially local perturbations acting on the Born-Oppenheimer states. 
The energy shifts are approximately proportional to the amplitude squared of the unperturbed states at the position of the avoided crossing.
In particular, the near-threshold states, which are spread out across several potential wells, already undergo some measure of quantum reflection as they encounter the increasingly deep potential wells between $1200 a_0 \le R \le 1500 a_0$.
This reduces their amplitude near the region of non-adiabatic coupling, wherefore their energy shifts are much smaller than those of the localized states bound closer to the avoided crossing. 

This behavior is clearly demonstrated by the states shown in \autoref{fig1:39}(a): the eigenstates with binding energies greater than $25\,$MHz and amplitudes concentrated around the narrow avoided crossing, experience measurable energy shifts compared to the states closer to threshold. 
Accordingly, in the stabilization diagram (\autoref{fig1:39}(b)) the shift can be noticed between the resonance states (appearing as flat plateaus) at lower energies. 
Moreover, all three calculations exhibit a different dependence on the inner boundary positions, resulting in different widths. 
This energy shift is also present in other electronic states of Rydberg molecules as well as in other Rydberg settings \cite{hummel_vibronic_2023, srikumar_nonadiabatic_2023,hollerith_quantum_2019}. 

Inspection of \autoref{fig1:39}(d) shows that the molecular lifetimes change significantly upon the inclusion of beyond-Born Oppenheimer effects, increasing by more than double. 
This lifetime depends on how strongly the wave function amplitude changes at the avoided crossing, which occurs both due to quantum reflection as well as non-adiabatic effects. 
These are illustrated by the large diagonal barrier in $\Lambda_{ii}$, which forces the wave function to tunnel through a classically forbidden region.
This leads to an additional exponential damping of the wave function's amplitude.
The same perturbative picture used previously shows that this exponential damping depends only on the height and width of the barrier, which remain nearly constant over the energy range of interest. 
Consequently, non-adiabatic couplings extend all molecular lifetimes by a similar factor which is nearly independent of the resonance energy, leading near-threshold states (at energies above $10\,$GHz) to possess lifetimes comparable to that of the $42s$ Rydberg atom, $41\,\mu \mathrm{s}$ \cite{branden_radiative_2009}.
These can be considered true bound states.

\begin{figure}[t]
    \centering
    \includegraphics[width= 0.47 \textwidth]{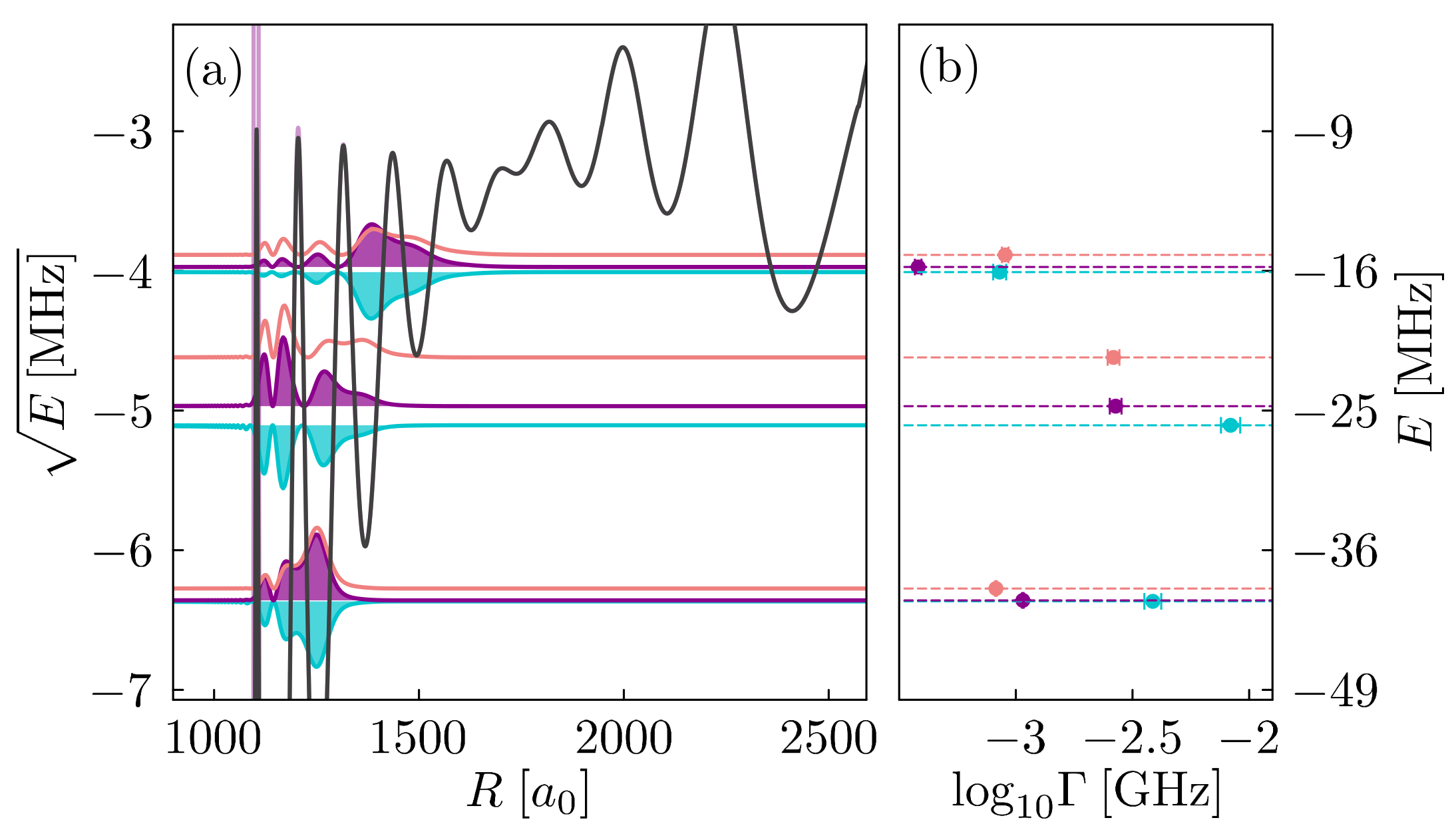}
    \caption{(a) Selected vibrational states of the adiabatic Rb$(39s)$+Rb$(5s)$ potential energy curve obtained using the Born-Oppenheimer approximation (turquoise), Born-Huang approximation (pink) and full coupled-channel (purple) calculation. The Born-Oppenheimer potential is given in black, while Born-Huang one is in light pink. 
    Energies are given with respect to the Rydberg $s$-level. (b) The widths $\Gamma$, shown on a logarithmic scale, of the corresponding vibrational states. The plot includes error bars obtained by comparing the Siegert and stabilization methods. }
    \label{36:mid-energies} 
\end{figure}

Not only do we find that the lifetimes are extended by non-adiabatic effects, but we even find a state that is predominantly bound by them. This can be clearly seen in the density of states calculated in the different approximations in \autoref{fig1:39}(c). In the Born-Oppenheimer approximation we find a prominent peak at $30\,$MHz with a broad red-detuned shoulder. The state associated with this shoulder narrows considerably and evolves into a clearly distinguishable peak when non-adiabatic couplings are included. 

The consequences of non-adiabatic coupling depends not only on the coupling strength, but also quite sensitively on the structure of the adiabatic potential curve. For example, the Born-Oppenheimer PEC for the $39s$+$5s$ molecule shown in \autoref{36:mid-energies}(a) possesses several wells of similar depth to the right of the butterfly drop. 
Here, even though the non-adiabatic couplings are similar in strength to the $42s$+$5s$ molecule considered above, the Born-Oppenheimer $42s$ PEC is intersected by the butterfly PEC just after a complete well forms. As a result, the more deeply bound vibrational states here are already more tightly confined by the Born-Oppenheimer PEC, but they tunnel through the narrow potential barriers to the right and  spread across a number of wells.
As in the previous case, we see a clear relationship between the spatial localization of the wave functions and the resulting energy shift -- the lowest and highest wave functions shown in \autoref{36:mid-energies} are more localized in the potential wells further away from the avoided crossing, leading to small non-adiabatic energy shifts. However, the wave function in the middle is more localized in the innermost potential well, concentrated close to the local perturbation.  This causes a more pronounced energy shift.

\begin{figure}[t!]
    \centering
    \includegraphics[width=0.45\textwidth]{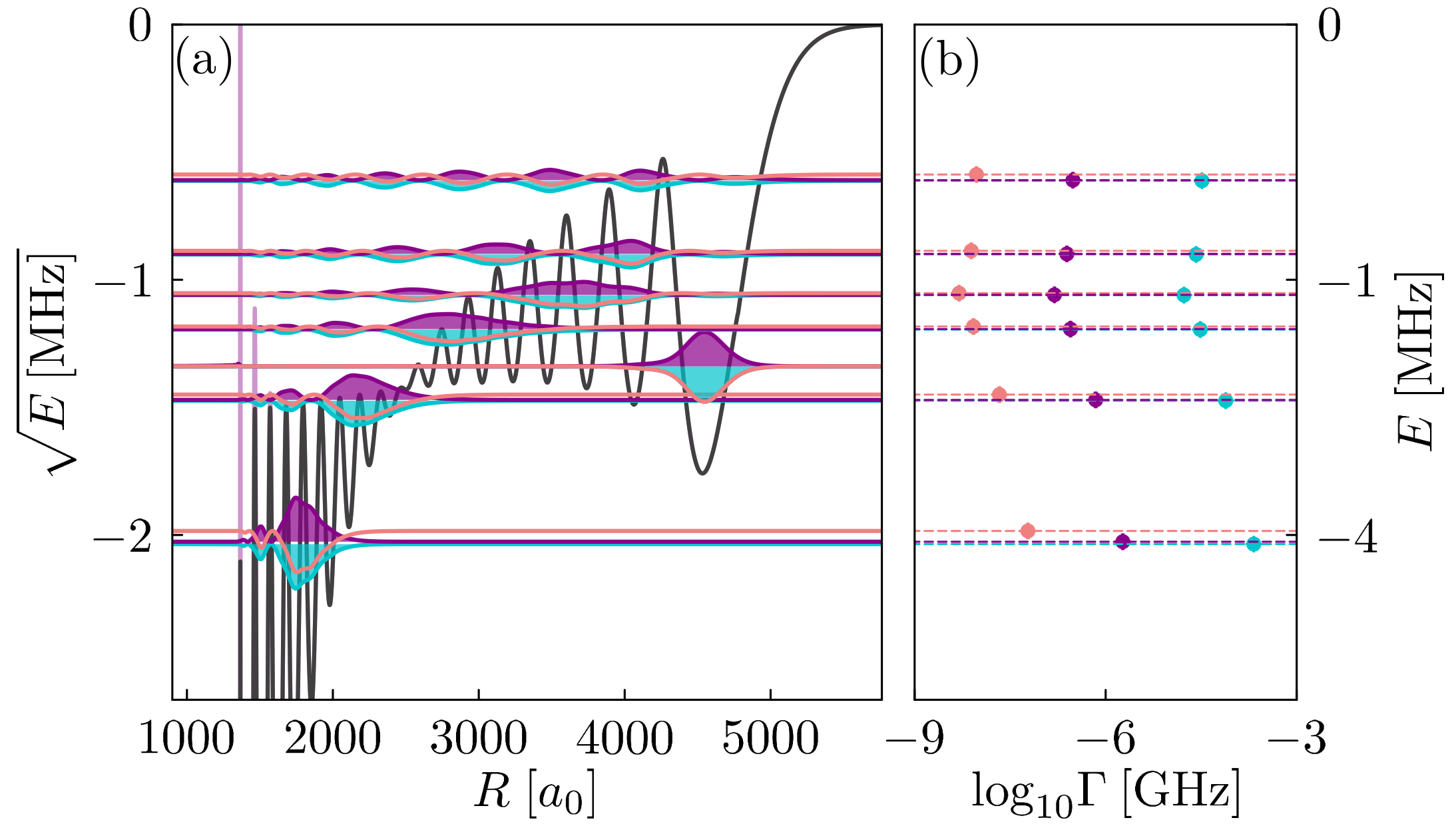}
    \caption{Selected vibrational states of the adiabatic Rb$(52s)$+Rb$(5s)$ potential energy curve obtained using Born-Oppenheimer (turquoise), Born-Huang (purple) and coupled-channel (pink) calculation. Born-Oppenheimer potential is given in black, while Born-Huang one is in light pink. Energies are given with respect to the Rydberg $s$-level. (b) The widths $\Gamma$, shown on a logarithmic scale, of the corresponding vibrational states. The plot includes error bars obtained by comparing the Siegert and stabilization methods.}
    \label{fig2:52}
\end{figure}

It is well-established that the Born-Huang approximation provides an upper bound for the vibrational ground state energy \cite{Book_born_huang}, while the Born–Oppenheimer approximation provides a lower bound \cite{epstein_groundstate_1966}. 
This additionally holds true for all states that we examined. In general, we find that the Born-Huang approximation strongly overestimates the effect of non-adiabatic coupling on the binding energies, which are typically closer to the Born-Oppenheimer prediction. 
On the other hand, the molecular lifetimes calculated in both approximations serve as a lower bound to the actual solution, with the Born-Huang approximation providing a more accurate estimate. 
In a naïve picture, the wave functions computed using the full coupled-channel approach are lower in energy than those calculated in the Born-Huang approximation, therefore they have to tunnel through a wider barrier, leading to longer lifetimes. 

The examples highlighted here paint a picture of the general behavior occurring in long-range Rydberg molecules: non-adiabatic couplings can significantly increase the lifetimes of molecular states, indicating that these couplings serve as an additional binding mechanism alongside quantum reflection. 
Depending on the strength of this additional coupling, Rydberg molecules can be stable against decay even to the extent that their lifetime approaches that of the bare Rydberg atom. 
We have selected these two Rydberg states because the adiabatic PECs approach one another closely enough for non-adiabatic effects to have an observable impact; this was the case for about 30\% of the Rydberg levels in the range $20<n<50$. 

Recent work has demonstrated that narrow avoided crossings in fact occur regularly in Rydberg molecules due to the unique binding mechanism and its direct dependence on the Rydberg electron's nodal structure \cite{eiles_katos_2024,hummel_synthetic_2021}.
In the present case of butterfly and $ns$ state non-adiabatic coupling, we can determine an approximate nodal condition following the argumentation of Ref.~\cite{hummel_synthetic_2021}. 
We find that the strength of the non-adiabatic coupling increases if the intersection between the diabatic potential curves (approximately determined by the condition $\delta_P[k(R,n)] = \mu_s$, where $\mu_s$ is the quantum defect of the Rydberg $ns$ state) occurs at a node of the the $ns$ wave function. 

\begin{figure}[t!]
    \centering
    \includegraphics[width=0.9\linewidth,]{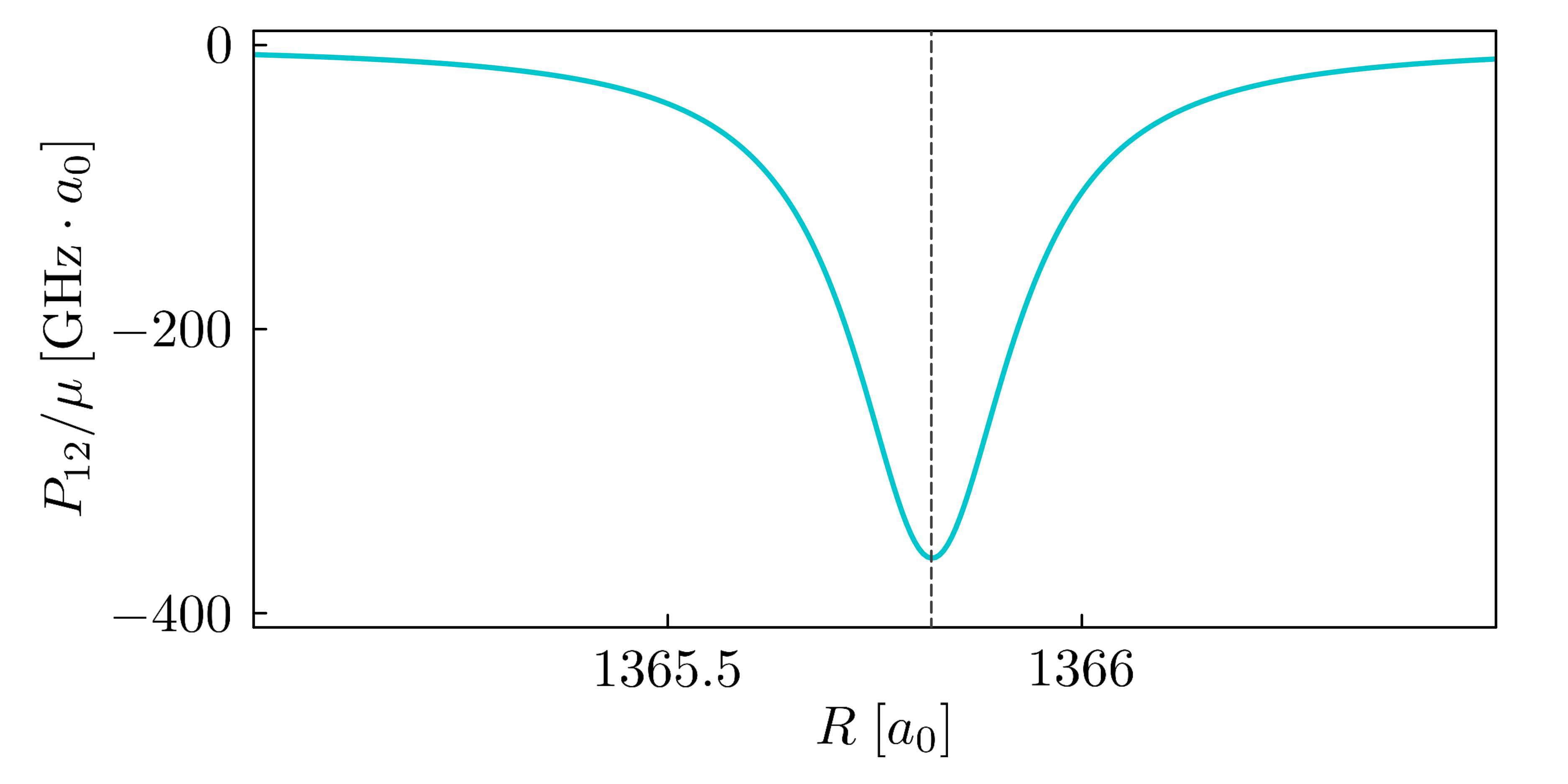}
    \caption{The off-diagonal non-adiabatic coupling matrix element P$_{12}$/$\mu$ for the Rb$(52s)$+Rb$(5s)$. The vertical dashed line indicates the position of the avoided crossing, which coincides with the maximum of $P_{12}$.}
    \label{fig:49_p}
\end{figure}

Indeed, we find such a case in the  $52s$+$5s$ Rydberg molecule. Here, a node of the $52s$ state lies less than five Bohr away from the crossing point.  
The adiabatic potentials are separated at the avoided crossing by a gap of only $\sim  10\,$MHz. 
Here, the diagonal non-adiabatic correction $\Lambda_{ii}$ generates a nearly singular peak atop the innermost well, as shown in \autoref{fig2:52}. 
Its magnitude of $\sim 800 \,\mathrm{GHz}$ significantly exceeds the scale of the potential curves. 
The off-diagonal matrix element is also sharply peaks, as shown in \autoref{fig:49_p}.

When the non-adiabatic coupling becomes so strong, neither of the approximations based on a single-channel adiabatic picture give even qualitatively accurate lifetimes. 
The single-channel lifetimes deviate from those obtained using the coupled channel equations by two to three orders of magnitude (see \autoref{fig2:52}(b)).
Not surprisingly, the Born Oppenheimer approximation underestimates the lifetime. 
In contrast to the general trends observed when the couplings are less singular, the Born-Huang approximation dramatically overestimates the lifetime, making it necessary to solve the full coupled-channel equations \cite{hummel_synthetic_2021, worth_beyond_2004, srikumar_nonadiabatic_2023}. This is further understood by considering the Landau-Zener transition probability
\begin{equation}
\label{eq:LZ}
    P_\text{LZ} = \exp\left(\frac{-2\pi \Delta^2}{4vs}\right),
\end{equation} 
where $\Delta$ is the spacing between the PECs $U_1$ and $U_2$ at the avoided crossing, $s$ is the differential slope of the crossing, and $v$ is the semiclassical velocity. 
$P_\mathrm{LZ}$ reaches a value of $0.999$ at threshold. 
This indicates that a diabatic single-channel picture for the level crossing would more accurately reproduce the physics. In the first two examples, the Landau-Zener transition probability at the crossing points does not exceed 50\%, and hence the adiabatic approximations are more accurate.

\section{Summary and Outlook}
We have shown that non-adiabatic couplings indeed provide an additional binding mechanism for long-range Rydberg molecules, stabilizing them against predissociation. 
Especially for extremely narrow avoided crossings, non-adiabatic effects have a very strong impact, becoming the dominant binding mechanism over quantum reflection.
The lifetime of the Rydberg molecule can be increased until it reaches the upper bound set by the lifetime of the Rydberg atom itself.

The presented examples were calculated for a specific set of Rb electronic phase shifts and a finite electronic basis.
Therefore, our results qualitatively describe the extent to which non-adiabatic coupling can effect the binding of Rb$(ns)$+Rb$(5s)$ molecules, but the specific $n$ levels where exceptional behavior occurs are not quantitatively guaranteed.
For example, a shift of $\sim1\,$MeV in the position of the $P$-wave shape resonance would shift the position of the nearly diabatic crossing to the $53s$ state.
Concrete predictions at this level of accuracy will require improved knowledge of the phase shifts, and can indeed be used to improve the procedure used to fit them to experimental data \cite{hummel_vibronic_2023}.

The effects of non-adiabatic couplings shown in this paper are somewhat weaker than might be generically expected. 
One reason for this is that the vibrational states already undergo quantum reflection off the multiple deep wells to the right of the avoided crossings.
This suppresses their amplitude near the coupling region. 
In contrast, the butterfly curve intersects the low-$\ell$ states at larger $R$ values in Rydberg molecules formed with a cesium ground-state atom, typically after just one or two potential wells have formed, which will strongly reduce the influence of this reflection. 
The effects of beyond-Born Oppenheimer physics which will also become more pronounced in Rydberg molecules with a smaller reduced mass, for example those formed out of potassium \cite{peper_heteronuclear_2021} or lithium \cite{schmid_rydberg_2018} atoms. 
Examining the potential energy curves obtained when spin-orbit and hyperfine spin degrees of freedom are included \cite{eiles_hamiltonian_2017}, one sees that sharp avoided crossings occur with increasing frequency as the multiplicity of the potential curves grows. 
As the number of potential curves and variability in energy level splittings increases when hyperfine and spin-orbit effects are included, it will only become much easier to approximately satisfy these conditions and find arbitrarily narrow avoided crossings.

\bibliography{MyLibrary} 

\end{document}